%% file: main.tex
\documentclass[manuscript, nonacm]{acmart}
\usepackage[utf8]{inputenc}
\usepackage{comment}
\usepackage{graphicx}
\usepackage{caption}
\usepackage{subcaption}
\usepackage{gensymb}
\usepackage{wrapfig}

\acmDOI{}
\acmYear{2026}

\begin{document}



\input{title-abstract.tex}

\maketitle

\input{body.tex}

\bibliographystyle{ACM-Reference-Format}
\bibliography{main}

\end{document}

%% file: title-abstract.tex
\title{Short-time, Wavelet-inspired Mouse Submovement Detection}

\author{Auejin Ham}
\author{Ben Boudaoud}
\affiliation {
    \institution{NVIDIA}
    \city{Durham}
    \country{USA}
}


\begin{abstract}
Submovements are ballistic components of human motion constituting a large part of motor interaction and arising from the cyclical and overlapping cognitive processes of perception, motor planning, and motor execution.
Extracting submovements is challenging as the motions tend to \emph{overlap}, or start before the previous ends.
We propose and evaluate use of a wavelet-inspired technique to accurately locate and parameterize submovements from one-dimensional speed time series.
Our method employs a self-weighted loss refinement step to identify and improve regions of poor quality of fit, a challenge for simpler wavelet transforms.
We demonstrate the accuracy of our method by presenting analysis of $\sim$6,400 1-2s trials of synthetic egocentric camera (first-person shooter) aim data for which we know ground truth, modeled from a similarly sized real data set of 13 users.
We compare our method to dual-threshold and the persistence 1D segmentation techniques and note challenges and opportunities for future improvements.
\end{abstract}

%% file: body.tex
\section{Introduction}

\begin{figure}
    \centering
    \includegraphics[width=0.8\columnwidth]{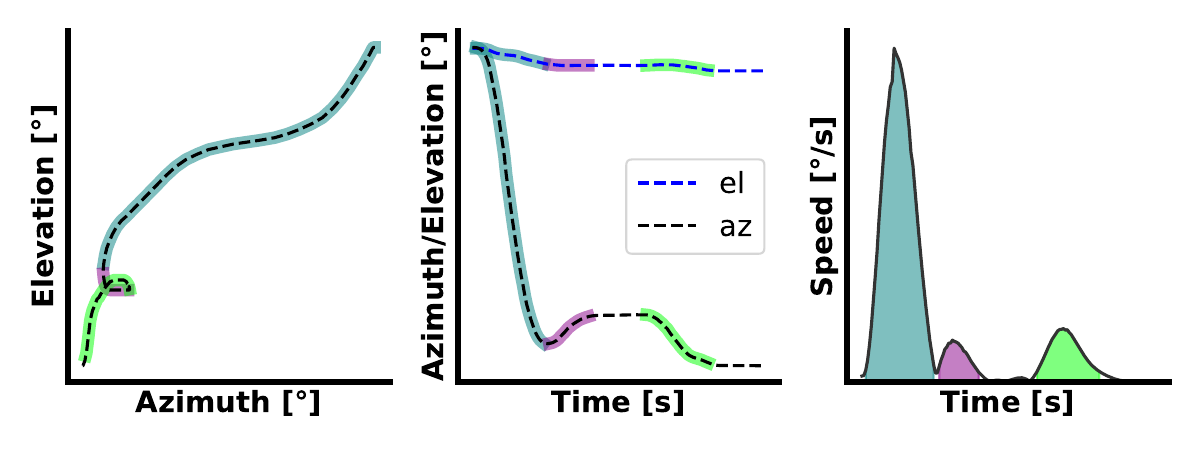}
    \vspace{-4mm}
    \caption{An example decomposition of a 3 submovement motion using a simple velocity-threshold based detector. (Left) the azimuth/elevation trace of a (2D) view direction ignoring time. (Center) a plot of azimuth (gray) and elevation (black) vs time showing the same submovements. (Right) the combined view speed vs time plot demonstrating ballistic characteristics.}
    \label{fig:example_decomp}
\end{figure}

Much of human motion in common contexts is composed of super-imposed discrete, pseudo-ballistic portions of movement referred to as \emph{submovements} \cite{hogan2012dynamic}.
Submovements have been demonstrated to occur in one \cite{meyer1988optimality, novak2000kinematic}, two \cite{fradet2008origins, huysmans2012submovement, chen2015structure}, and three \cite{chen2012submovement, liao2012predicting} dimensional tasks, and generally share a common shape in the velocity magnitude (i.e., speed) domain.
An example decomposition of these submovements, taken from a real, two dimensional (pitch/yaw mapped directly to mouse motion), first-person shooter (FPS) aim data set is provided in Figure \ref{fig:example_decomp}.

By better understanding submovement speed, count, and error we gain insight into how humans perform various motor tasks, and how these tasks in turn, shape our motor planning and control strategies as humans.
Submovement properties have already been demonstrated to correlate to a number of different human factors, and are a useful tool for explaining limited or increased performance of certain populations in various computerized tasks \cite{dipietro2009submovement, walker1997age, rohrer2004stroke, hwang2003mouse}.
However, in spite of availability of accepted models for individual submovements' speed profiles and value demonstrated in studying the \emph{overlap} of nearby submovements, there are no broadly published, computationally efficient methods for decomposing submovements assuming a single unified model that fits all individual submovements.

Wavelet decomposition is a popular technique in which a \emph{mother wavelet} is scaled in amplitude and dilated in width in order to model a series as a composition of, assumed orthogonal, constituent components.
The technique is a near ideal fit for detection of overlapping submovements as the problem formulation lends itself to an iterative decomposition of individual movements of known shape (but unknown amplitude, width, and location) from a greater time series with unknown structure.
In this work we describe a technique for not just applying a wavelet-inspired transform to speed time series for (potentially overlapping) submovement decomposition, but also a loss metric and regression strategy to refine this selection.
We use our approach to model a data set of experienced PC gamers aiming in an FPS environment and demonstrate significant differences when compared to more traditional non-overlapping segmentation techniques on real data.

\section{Background}
It has long been observed that human motion is driven by an intermittent motor control strategy allowing discrete, overlapping \emph{submovements} to occur as part of greater motion \cite{gross2002neural, novak2002use}.
Underlying this intermittent control strategy is a biomechanical speed-error relationship, implying that the faster we move, the larger the errors we make in directing motion, resulting in an optimal control strategy that uses a combination of large, fast movements and smaller, slower refining motion(s) \cite{meyer1988optimality, dounskaia2005influence, fishbach2007deciding}.
Some have theorized that the popular Fitts' law \cite{fitts1954information} arises as a result of the nature of iterative control in a speed-error constrained space \cite{van1995fitts, crossman1983feedback}.
Adaptive model theory (AMT) makes the case for human motion as an optimal intermittent motor control strategy, adapted to task-dependent non-linear interactions \cite{neilson2005AMT}.
More recently this intermittent control strategy has been formalized into motor predictors such as the BUMP model \cite{bye2008bump}.
Inherent in this work, is the idea that overlapped intermittent motor control gives rise to the overlapped submovements or \emph{online corrections} observed in prior art \cite{gomi2008implicit}.

By decomposing greater human motion into constituent submovements we gain insight into how both the human and task impact motor control strategy.
Prior work demonstrates that submovements become larger/fewer/more overlapped, and thus more time-efficient during stroke recovery \cite{rohrer2004stroke} and slower/more separated in Parkinson's patients \cite{dounskaia2009submovements}.
Similarly, age has been demonstrated to play a significant role in submovement strategy, with not just speed and count of submovements changing, but also their composed structure \cite{walker1997age}.
Analysis of mouse usage in first-person targeting tasks demonstrates additional submovements incur super-linear task completion times, driving elite esports athletes to optimize ease of performing consistent, repeatable submovements \cite{boudaoud2022mouse}.

\subsection{Existing Decomposition Strategies} \label{sec:decompmethods}
A number of algorithms already exist for decomposing submovements from a greater motion times series.
Meyer's original work, and subsequent follow-up studies, proposed a fixed 2-submovement model for a one dimensional, knob-based, rotational task, with a velocity threshold applied to delineate the first submovement from the second \cite{meyer1988optimality, novak2000kinematic}.
It is worth noting that in many, reasonably simple tasks, a fixed 2-submovement model (primary and corrective) works well enough to describe targeting processes \cite{hoffmann2016fitts}.
Later work used combined amplitude-time criteria to pair corrective motions that occur in close proximity to large primary movements \cite{fradet2008origins, walker1997program}.
However, all of this work delineates \emph{types} of submovements \cite{wisleder2007role}, particularly \emph{primary} motions which occur quickly to move towards a target, and optional secondary \emph{corrective} motions which occur more slowly to refine aim when the primary motion does not reach its intended target.

As tasks grow more spatially and temporally challenging the presence of additional corrective submovements is observed.
The techniques above can be extended to multiple submovements through use of a simple start-stop (dual) threshold technique in which a new submovement is declared whenever (1) the prior submovement has ended (i.e., velocity has been below the end threshold) and (2) the velocity crosses the start threshold \cite{boudaoud2022mouse, walker1997program}.
In at least one improved algorithm, the end point position of the movement is also considered in determining the end point of a submovement \cite{naghibi2017modified}.
These hysteresis-based detectors work well when describing initialization, pause and verification times, the periods of little to no motion before, during, and after task motion occurs, and can segment arbitrary counts of primary and corrective movements but have two distinct downsides:

\begin{itemize}
    \item They cannot detect \emph{overlap} between submovements (any one time point belongs to a single submovement)
    \item When submovements do (heavily) overlap they are classified as a \emph{single submovement} with an alternate shape
\end{itemize}

Prior work provides models for the speed profile of individual submovements, specifically Plamondon et al. find that a support bounded log-normal fit their data set best when compared using MSE \cite{plamondon1993modelling}.

Some work has utilized these models for iterative fitting decomposition strategies that do not rely on threshold-based detection \cite{rohrer2003avoiding, rohrer2006avoiding, gowda2015accelerating, krishnan2017segmenting}.
While these strategies use iteratively optimized fitting approaches to attempt to solve the non-linear optimization problem of submovement fitting under various constraints, we are the first to propose the use of an efficient wavelet transform to provide valid initial guesses for decomposition.

\subsection{Submovement Overlap}

\begin{wrapfigure}{r}{0.5\textwidth}
    \includegraphics[width=0.45\textwidth]{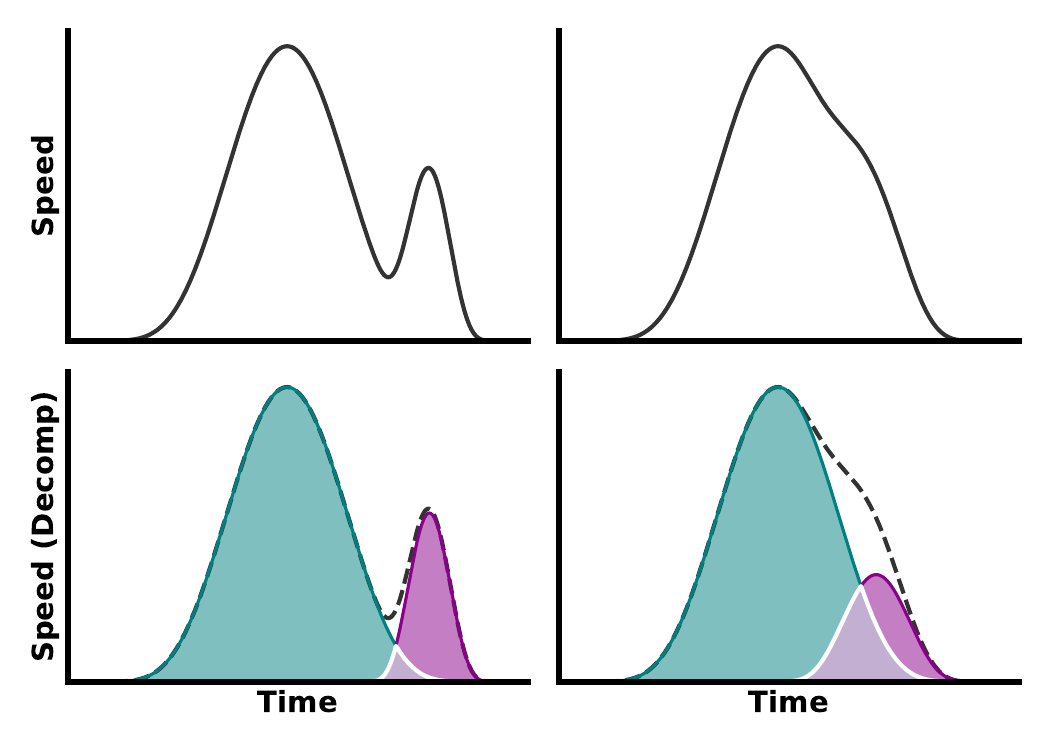}
    \caption{Demonstration of overlap of two different submovements. (Left) a small overlap region (shaded in white) with two distinct submovements clearly resolve-able. (Right) a larger overlap makes it more difficult to segment individual submovements.}
    \label{fig:overlap}
\end{wrapfigure}

The \emph{overlap} of submovements is an interesting area of study, we define this overlap as the amount two (or more) submovements are executed simultaneously (superimposed) by the motor system.
Overlap can be measured either in time or as a percentage of the net displacement of the movements.
Figure \ref{fig:overlap} provides an example of a low and high overlap speed profile for two submovements.
Prior art has noted that the overlap of submovements is a primary indicator of \emph{smoothness} (i.e., lack of jerk) in overall motion, and can serve as a good indicator of neurological improvement following traumatic events \cite{rohrer2004submovement}. 
However relying on aggregation of jerk, which can be significantly impacted by amplification of noise in the 3rd derivative of motion, as a proxy for overlap is undesirable compared to more direct measurement of the overlap time/displacement of constituent movements.

\subsection{Emerging Applications}
Historically submovements were primarily studied in the clinical context \cite{hwang2003mouse,dounskaia2009submovements,rohrer2004stroke}, but more recent work poses the value of submovement analysis to a wider scope including esports \cite{toth2023exploring, boudaoud2022mouse} and robotics \cite{lockwood2022leveraging}.
Other work explores modeling the count and location of submovements probabilistically, based on task index of difficulty \cite{chen2015structure}, and in-turn attempts to use characteristics of submovements to improve pointer assistance techniques \cite{hourcade2010pointassist, lee2020autogain}.


\section{Algorithm}

\subsection{Submovement Speed Model}
\begin{figure}
    \centering
    \includegraphics[width=\columnwidth]{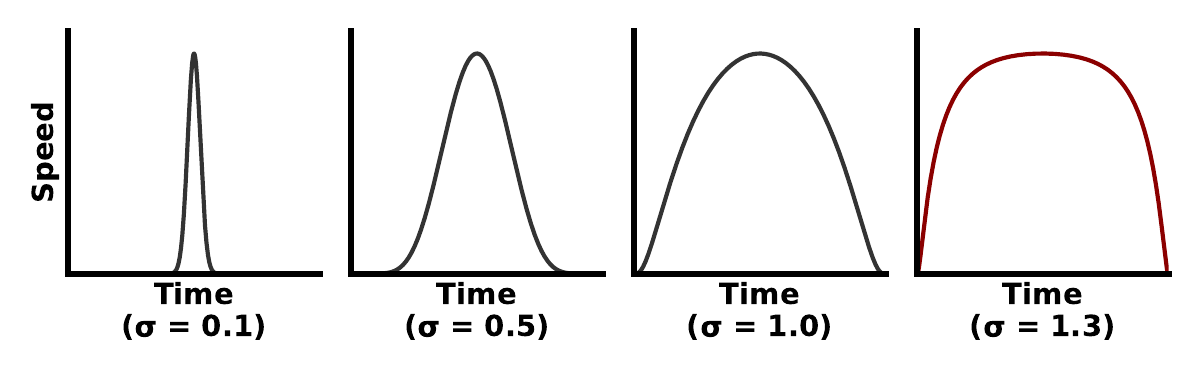}
    \vspace{-8mm}
    \caption{Plot of our mother (speed) "wavelet" with dilation parameter ($\sigma$) ranging from 0.1-1.3. We (typically) only consider values of $\sigma \in [0,1]$ as valid dilations. (Far right) values of $\sigma$ above 1 result in flattening the top of the waveform and eventually produce bimodal results which could model multiple submovements.}
    \label{fig:motherwavelet}
    \vspace{-6mm}
\end{figure}
We use a modified version of the support bounded log-normal model proposed by Plamodon et al. \cite{plamondon1993modelling} for our submovement speed model, or mother "wavelet", as provided in Equation \ref{eq:motherwavelet} and visualized in Figure \ref{fig:motherwavelet}, where $t$ is the current time in [$T_0$, $T_1$], the submovement start and end time respectively.
The $\sigma$ term is the width parameter we will use for dilations in the wavelet transform.
It is worth noting that this technically invalidates our approach as a wavelet transform as the model is not orthogonal under dilations (i.e., choice of different $\sigma$ values do not form a basis the way dilations of a sinc/sinusoid, Gabor, or Haar mother wavelet do).
No choice of valid submovement model from Plamodon's work would meet this requirement, so we simply move forward based on the best possible fit.

\begin{equation}
\label{eq:motherwavelet}
\psi_{T_0, T_1, \sigma}(t) = \frac{T_1-T_0}{\sigma\sqrt{2\pi}(t-T_0)(T_1-t)}e^{\frac{-1}{2\sigma^2}ln(\frac{t-T_0}{T_1-t})^2}
\vspace{2mm}
\end{equation}

This model did a good job of fitting nearly all single-submovement data we had available, suggesting the need for fitting skew is minimal.
In fact the use of a skew parameter may increase the likelihood  of confounding two highly overlapped submovements with a single submovement as demonstrated in the right side of Figure \ref{fig:overlap}.
It is worth noting that this model is essentially a smoothed version of the constant acceleration (ballistic), or zero-jerk, model of human motion, agreeing with prior art stating constituent human motions are as locally smooth as possible, or minimize acceleration. 

\subsection{Decomposition Algorithm}
\label{sec:decomposition}

\begin{figure}
    \centering
    \includegraphics[width=0.6\textwidth]{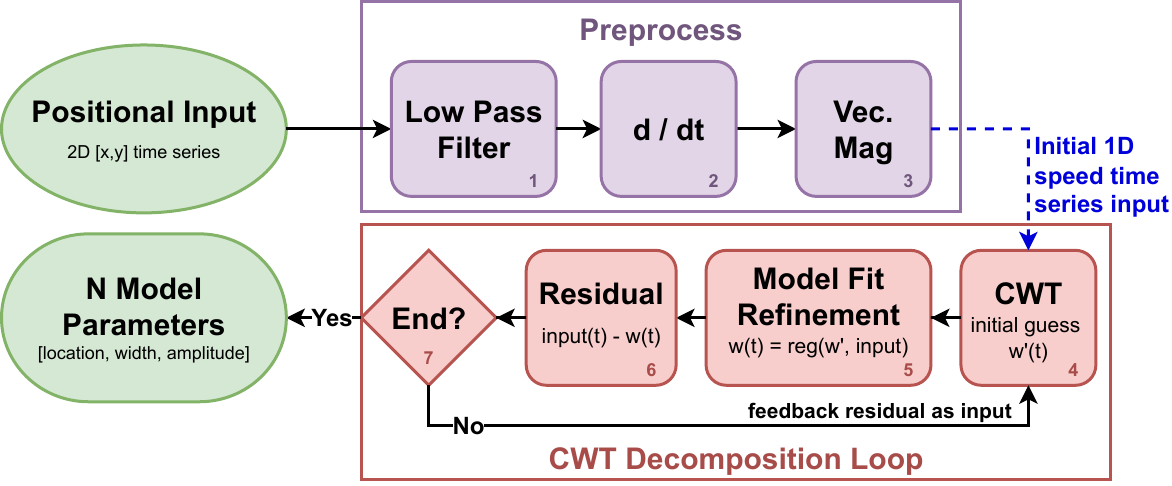}
    \caption{High-level flow of our decomposition algorithm. Pre-processing entails: (1) low-pass filtering of the 2D time series, (2) taking a derivative to a 2D velocity time series,  (3) a vector magnitude to create a 1D speed time series. This 1D time series is then (iteratively) decomposed by our decomposition loop as follows: (4) a CWT provides an initial guess for (5) refinement of model parameters using linear regression on our self-weighted loss function, (6) we compute a residual by subtracting the fit model from the input time series and (7) determine whether to stop iterating based on comparison of the model to this residual. When not at the end condition, we feedback the remaining residual as the next input to our CWT decomposition loop.}
    \label{fig:DecompFlow}
\end{figure}

\subsubsection{Sampling and Preprocessing}

We suggest sampling movement data at rates well (>10x) above Nyquist rates for decomposition, as the signal sampling rate provides resolution for both localizing submovements and evaluating model fit accurately.
We begin by low-pass filtering (0th order/positional) motion data at 7-11 Hz using a 5th order Butterworth low-pass filter.
This reduces sharp features, such as measurement artifacts, which might otherwise cause wavelet decomposition to fit noise or confound quantization effects with delays in motion.
Next we perform a first derivative over this (multi-dimensional) position and take a vector magnitude to create a one dimensional speed time series.

\subsubsection{CWT Decomposition Loop}

\begin{wrapfigure}{r}{0.5\columnwidth}
    \includegraphics[width=0.48\columnwidth]{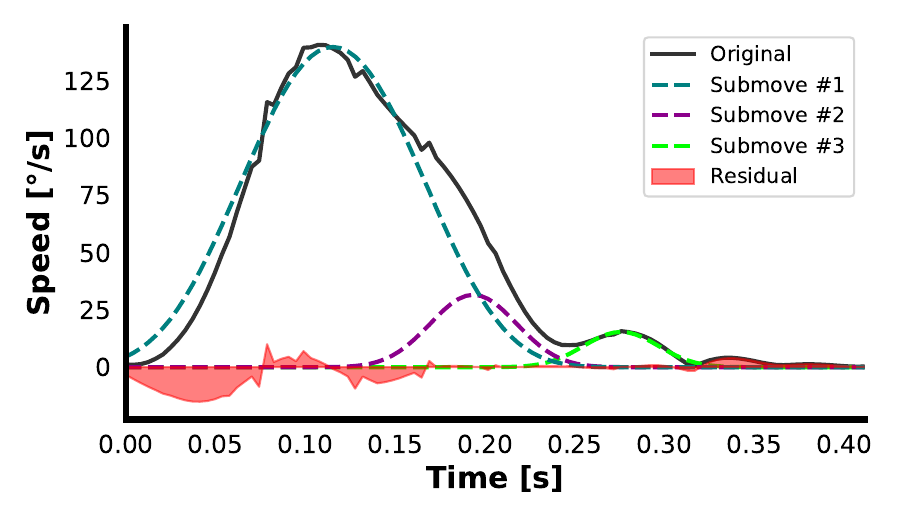}
    \caption{Example of a motion time series w/ 3 fit submovements, residual error is shaded in red. Note how misalignment or misprediction of submovement width/location creates negative residuals, in this case about 0.05 and 0.15 s into the trial. The remaining unfit submovements (around 0.1 and 0.33 s in the trial) are shown as positive residual.}
    \label{fig:residual}
\end{wrapfigure}

We use the SciPy signal processing library's CWT method (\texttt{scipy.signal.cwt}) \cite{2020SciPy-NMeth} for our implementation; however, this is not a requirement, and a custom/optimized wavelet decomposition method could be substituted.
We (naively) decompose the speed time series using a CWT with our submovement model as the mother wavelet, finding candidate locations and widths using the \texttt{argmax} over the resulting CWT amplitude matrix.
We treat this location, width, and amplitude as an initial guess for a refinement step using a dynamic programming model.
We then perturb these 3 parameters of the fit model using a weighted loss further described in Section \ref{sec:weightedloss}.
This refinement improves our initially discovered parameters to those better fitting the time series around speed peaks.


We obtain a residual by subtracting the final fit model from the corresponding input speed time series, and continue decomposing submovements from this residual until satisfying the termination criteria. 
We only use the positive-going portion of the residual for additional modeling, avoiding the issue of wavelet decomposition fitting negative-going residuals caused by misalignment or misprediction of model parameters, as demonstrated in Figure \ref{fig:residual}.

\subsubsection{Parameter Refinement}
\label{sec:weightedloss}

Rather than rely solely on (R)MSE metrics to evaluate goodness of fit of the model above to time series, we propose a self or \emph{wavelet-weighted} metric as an additional method for measuring fit quality.
Unlike (R)MSE which weights all points of fit equally, and can compensate large(r) mismatch around peaks with smaller mismatch around the low valued tails, this method up-weights regions which contribute most to displacement (i.e., those of greatest speed).
This is accomplished by multiplying/weighting the residual error in a region (i.e., sum after subtracting the fit model) by the peak-normalized wavelet model (plus an offset) over this window as demonstrated in Equation \ref{eq:WeightedLoss}.

\begin{equation}
\label{eq:WeightedLoss}
    Loss = \frac{1}{D}\sum_t{\left(v(t) - \psi(t)\right)\left({\frac{\psi(t)}{\max(\psi)} + C}\right)}
\vspace{2mm}
\end{equation}

In the equation above $v(t)$ is the input speed time series and $\psi(t)$ is the fit submovement speed model, with their difference ($v(t)-\psi(t)$) representing the residual loss over time.
$D$ is the total displacement of the movement (the integral of $v(t)$ over the time window in question) and serves to normalize results.
By dividing $\psi(t)$ by its maximum value, again over the time window, we additionally normalize the peak of the submovement to a value of 1, with the rest of the model obtaining values < 1. 
$C$ is an arbitrary offset, which we choose fix to 0.3, ensuring that regions where the speed model goes to 0 are still weighted as part of the fit.

\begin{figure}
    \centering
    \includegraphics[width=\columnwidth]{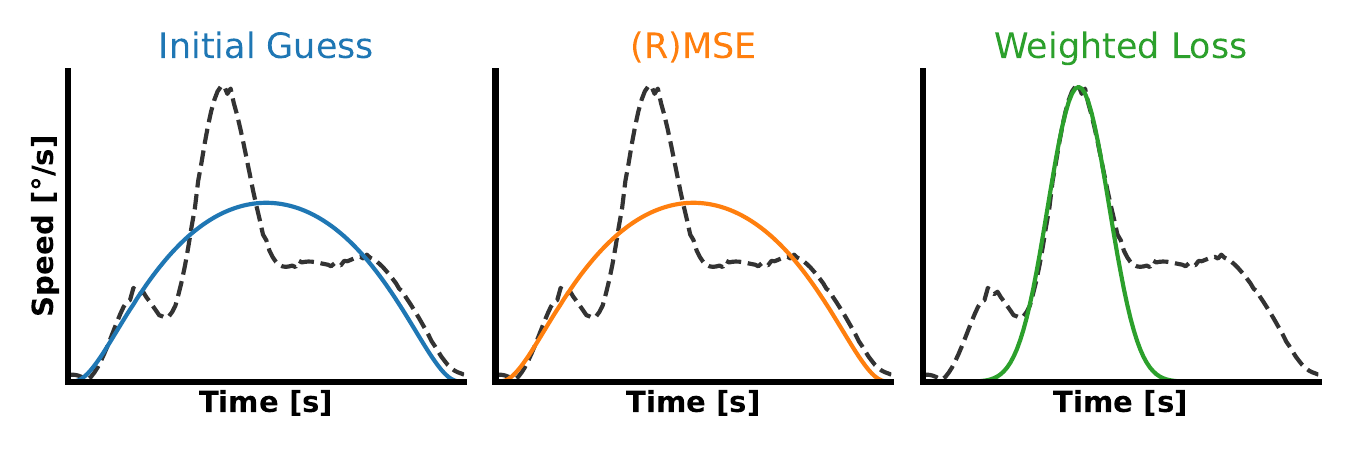}
    \caption{Demonstration of refinement of a poor fit based on (R)MSE and self-weighted loss. The (left) initial guess is based on a raw CWT result. We then linearly regress model parameters based on (center) (R)MSE loss, producing a result very similar to the initial guess or (right) self-weighted loss, correctly identifying a single peak.}
    \label{fig:refinement_after_cwt}   
\end{figure}

The primary goal of using this wavelet-weighted loss metric is that it captures cases where a single wavelet is over-fit (typically in width) to multiple submovements, as demonstrated in Figure \ref{fig:refinement_after_cwt}.
In these cases, MSE is often minimized by performing a good fit at the start/end of the model at the cost of a poor fit during its peak speed.

\begin{figure*}
    \centering    
    \includegraphics[width=\textwidth]{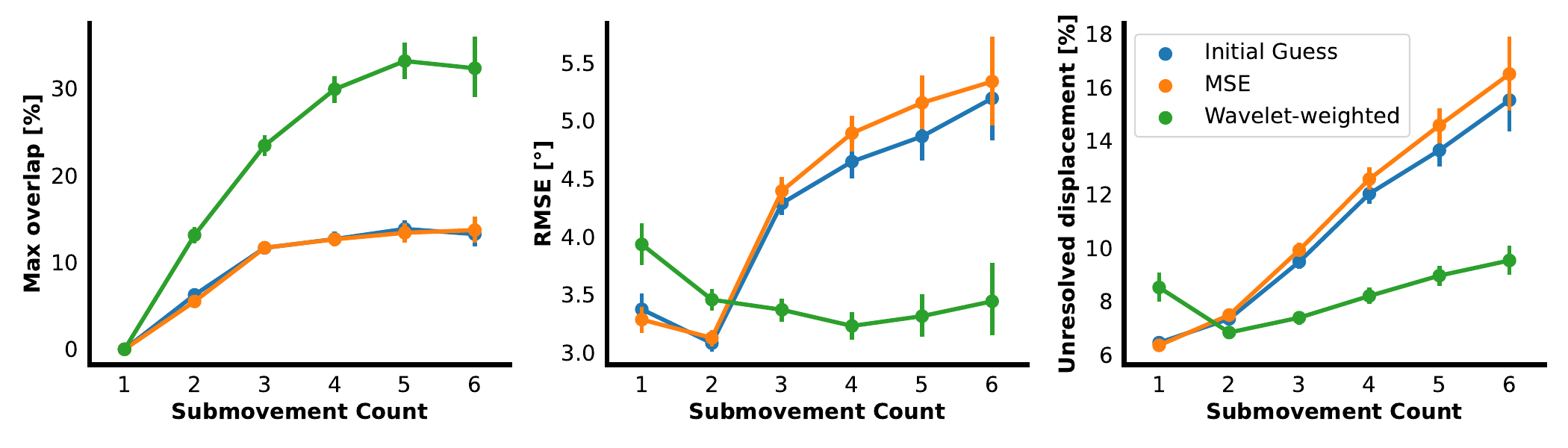}
    \vspace{-8mm}
    \caption{Evaluating submovement decomposition performance using real subject data. (Left) submovements tend to overlap more for trials with more submovements, wavelet-weighted loss more accurately reports this trend. (Center) wavelet weighted loss yields smaller RMSE and (right) less unresolved displacement as more submovements are found in a given trial. All error bars represent standard error metric.}
    \label{fig:realdata_result}
\end{figure*}

We used real data from a first-person shooter aim experiment (see Section \ref{sec:realdata} below) to validate our choice of this refinement criteria, as demonstrated in Figure \ref{fig:realdata_result}.
The wavelet-weighted refinement resulted in significantly higher maximum overlap as well as lower RMSE and unresolved displacement when compared to the CWT initial guess or (R)MSE-based refinement, particularly when the submovement count exceeds 2.
This is likely due to our self-weighted loss doing a better job of localizing individual peaks when multiple submovements are present, with the initial wavelet decomposition doing a better job of fitting smaller numbers of more distributed peaks based solely on RMSE.

\subsubsection{Termination Condition}
\begin{figure}
    \centering    
    \includegraphics[width=\textwidth]{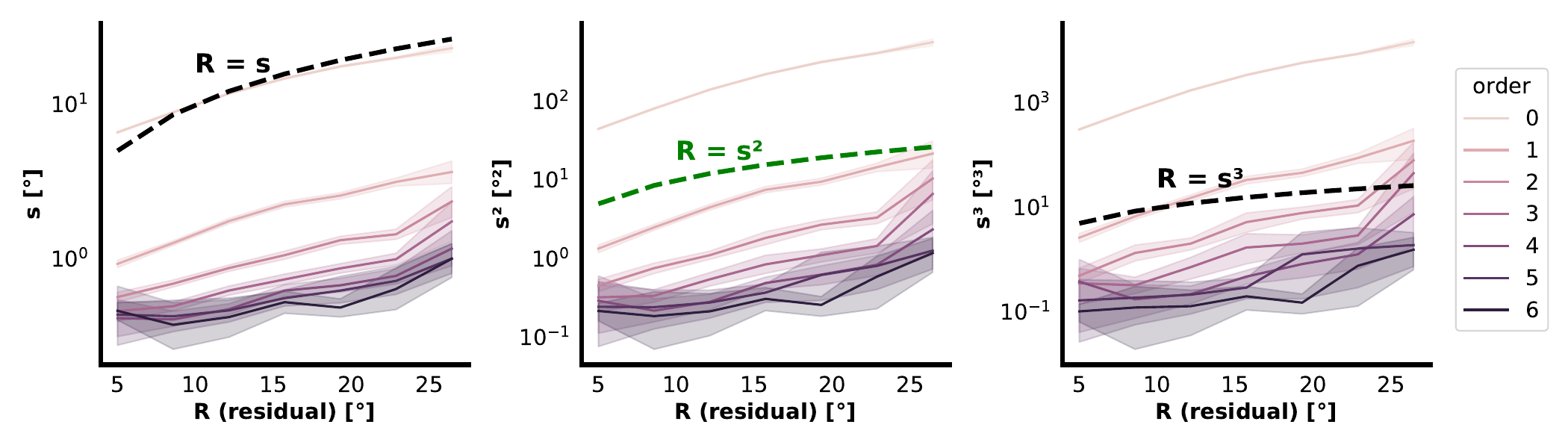}
    \vspace{-8mm}
    \caption{Comparing three termination conditions for submovement decomposition. $R$ is the residual, $s$ is the displacement of a fit submovement and the darkening pink lines represent the next 7 submovements at various decomposition steps. All y-axes are logarithmic. The center figure with a terminating condition of $R = s^2$ shows better separation of dominant (1st) and non-dominant (2nd-7th) predicted wavelets from a given  decomposition step.}
    \label{fig:terminating_condition}
\end{figure}
Our preferred termination criteria is defined as $s^2 \leq R$ where $s$ is the displacement of the currently fit submovement, and $R$ is the displacement of residual mouse movement at the current decomposition step. 
This portion describes whether the estimated submovement yields a displacement significant enough relative to the residual to justify continued decomposition. 
The reason behind using the square of $s$ is illustrated in Figure~\ref{fig:terminating_condition}. 
We produce all three figures by performing self-weighted decomposition on real mouse movement data. 
A good termination condition separates the dominant/current wavelet (order 1) and non-dominate wavelets (orders 2-7) from a given residual speed waveform at each decomposition step.



\section{Evaluation}

\subsection{Real Data}
\label{sec:realdata}

We collected $\sim$6500 trials of data from 13 users in an open-source FPS experimentation platform \cite{boudaoud2022firstpersonscience}.
This data was captured as part of a previous project, and we did not (initially) exclude any samples as all time series represented valid aim movements, though subjects did not necessarily successfully complete the FPS aiming task in all examples.
While this data represents valid aim examples, it does not include any ground truth labels for submovements, and as such cannot be used for absolute evaluation of our algorithm.

\begin{figure}
    \centering
    \includegraphics[width=\columnwidth]{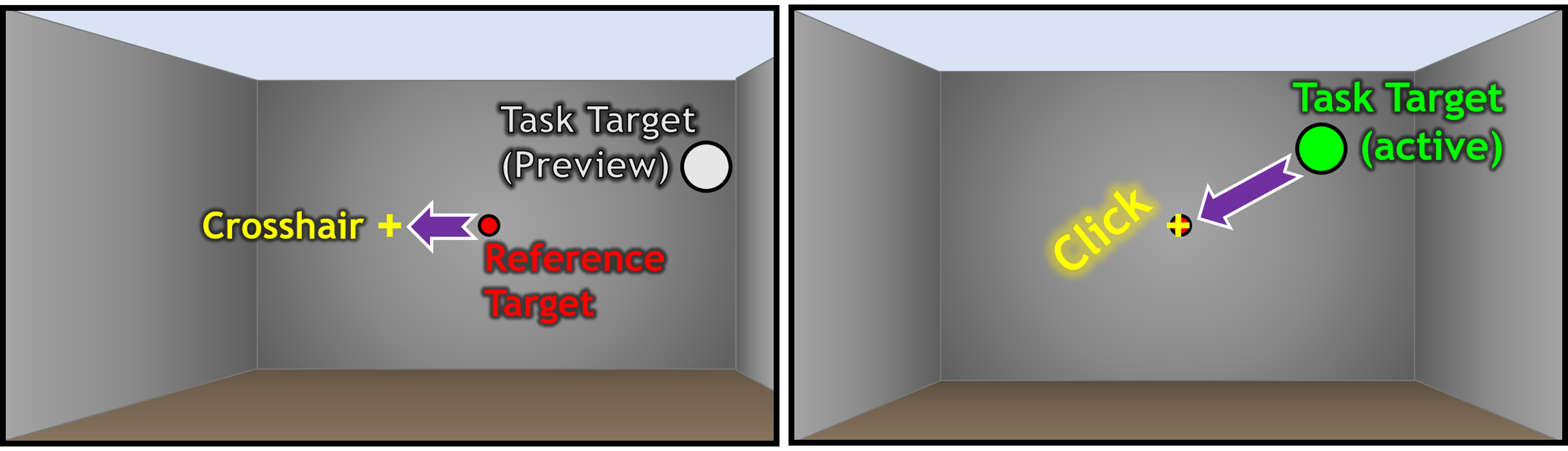}
    \caption{The task subjects completed in our experimental data. Before a trial (after a previous trial) the subject rotates the view to align at a red reference target with the always centered crosshair (top) while being presented a gray task target preview. Once the reference target is destroyed by clicking (bottom) subjects were instructed to rotate the view to aim and click on the (now green) task target as quickly as possible.}
    \label{fig:ExpTask}
\end{figure}

\subsubsection{Experiment Design}
Each subject was seating about 16 inches (400 mm) from a 240 Hz 1080p 25 inch LCD display. 
They used a Logitech G203 gaming mouse set to a 3200 DPI sensor resolution and 1 kHz polling rate. 
The data was collected with a purely linear translation-to-rotation mouse gain and no pointer acceleration (i.e., enhanced pointer precision) meaning in-game rotation was directly correlated to player mouse motion.
The experiment platform ran on an Intel Core i7-9700k @ 3.6 GHz, 32 GB of RAM, and an NVIDIA RTX 2080 Ti. 

In the experiment subjects completed a first-person targeting tasks aiming at stationary targets by moving the mouse to rotate the view direction and clicking to fire at, and if it hit, destroy the targets per the flow in Figure \ref{fig:ExpTask}.
Prior to each trial the subject aimed at a red reference target, returning their aim to a similar initial view direction.
While aiming at the reference target the task target was \emph{previewed} as a gray, inactive version of itself at its final size and position.
After the subject clicks with their crosshair over the red reference target, destroying it, the trial begins and the task target turns green.
Subjects then tried to move their aim as quickly as possible to the green task target and destroy it, with their score determined by how quickly they completed this action.

\subsection{Synthetic Data}
\label{sec:syntheticdata}


We used our real view direction data to create a statistical model that produces submovements of known parameters representative of real human input for evaluation.
While the resulting modeled series are not guaranteed to accurately reproduce actual human motion or the noise and quantization effects inherent to its measurement, they provide a ground truth for absolute evaluation of the algorithm.
This model allows us to perform studies of robustness of our algorithm to particularly challenging decompositions without relying on collecting and inspecting impractically large real data sets to find the few examples of these conditions. 

\subsubsection{Submovement Modeling and Generation}

\begin{figure}
    \centering
    \includegraphics[width=\columnwidth]{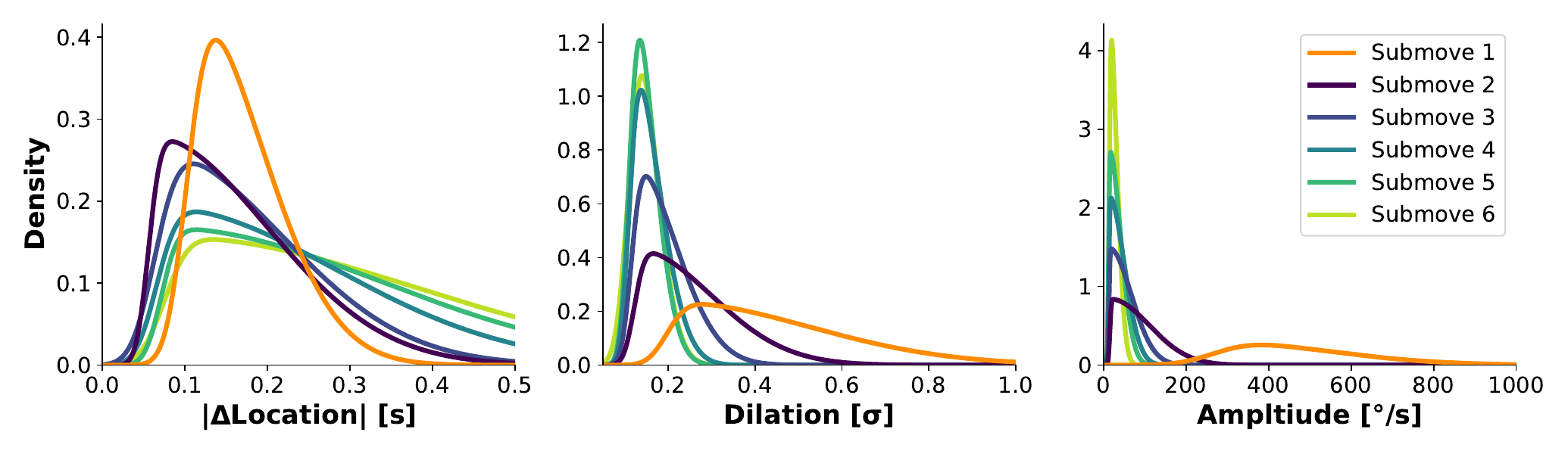}
    \vspace{-8mm}
    \caption{Probabilistic distributions of wavelet parameters: change in peak location ($\Delta$Location), dilation ($\sigma$), and amplitude ($A$) across different submovement counts. These distributions are sampled from to create our synthetic ground-truth trial data.}
    \label{fig:generatorparams}
\end{figure}

We used the real data from Section \ref{sec:realdata} to produce a statistical model for generating realistic mouse speed time series. 
We capture distributions of wavelet amplitude ($A$), dilation ($\sigma$) and the peak location ($0.5T_0 + 0.5T_1$) for each submovement in a trial.
Note that each trial was restricted to six or less submovements as less than 1\% of trials had 7 or more submovements in the real data.
These sample distributions are then fit with log-normal PDFs  (see Figure~\ref{fig:generatorparams}). 

\begin{figure*}
    \centering
    \begin{subfigure}[b]{.33\textwidth}
        \includegraphics[width=\textwidth]{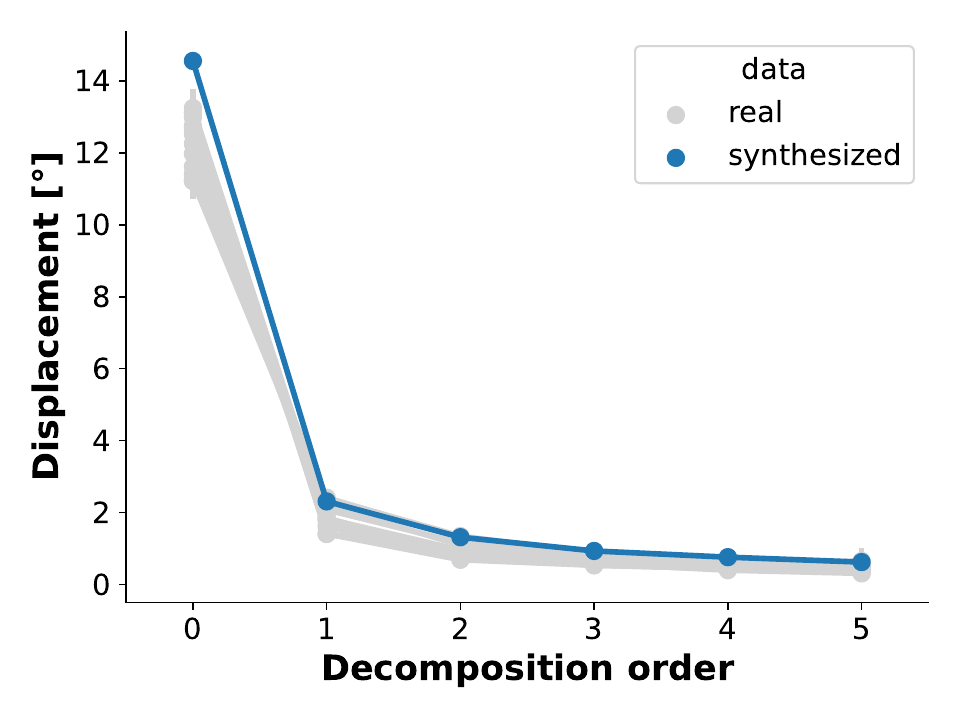}
    \end{subfigure}
    \hfill
    \begin{subfigure}[b]{.33\textwidth}
        \includegraphics[width=\textwidth]{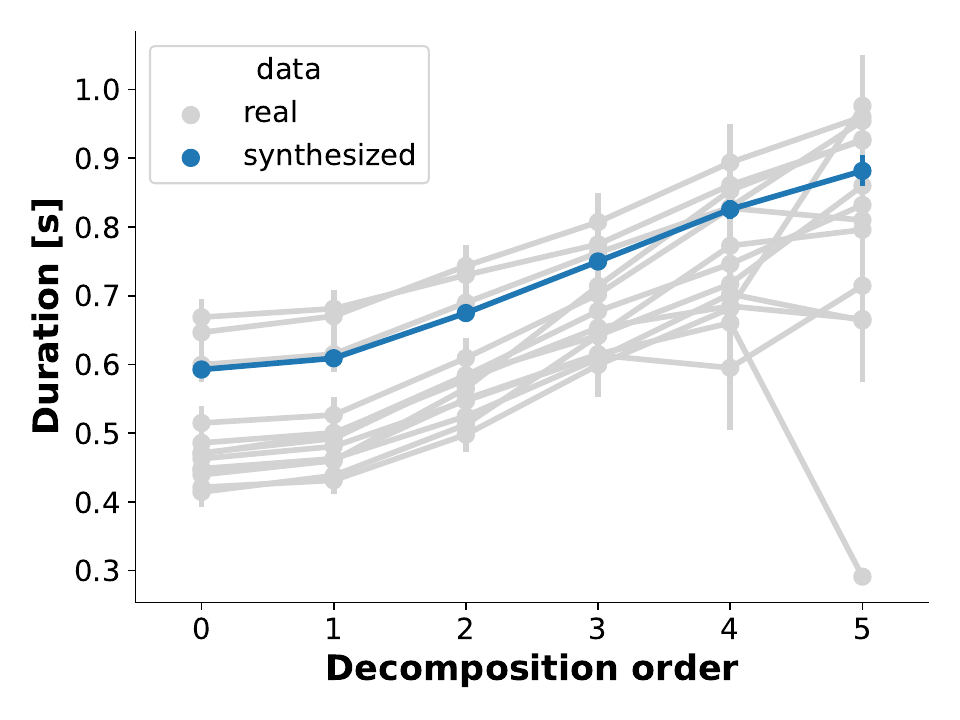}
    \end{subfigure}
    \hfill
    \begin{subfigure}[b]{.33\textwidth}
        \includegraphics[width=\textwidth]{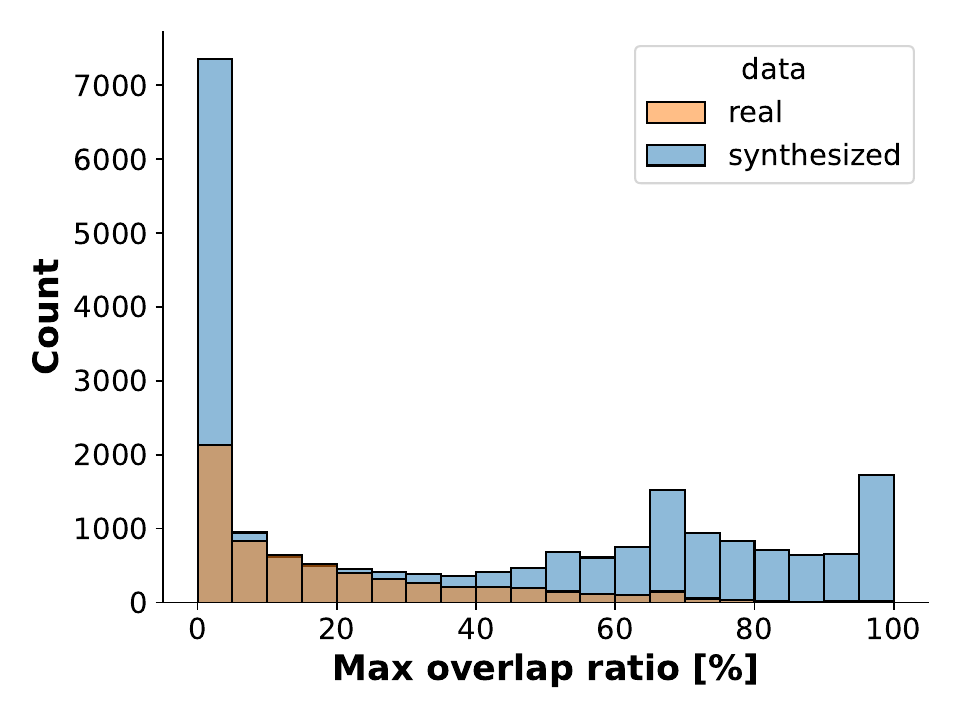}
    \end{subfigure}    
    \vspace{-6mm}
    \caption{Evaluation of synthesized trials (blue) as compared with real data from 13 participants (grey/orange). (Left) a comparison of total per-trial \emph{displacement} over submovement count. (Center) a comparison of duration of each submovement over its count. (Right) a histogram of the maximum overlap ratio across trials. Error bars on left and center figures represent standard error.}
    \label{fig:synthesized_trials}

\end{figure*}


We focus on demonstrating good detection with representative overlap/distance between motions, width ($\sigma$), and amplitude (A) parameters.
To achieve this goal, we over-generate trials and use a rejection sampling technique to match our overlap ratio distribution to that of the real data (see the rightmost plot in Figure \ref{fig:synthesized_trials}).
We synthesized 21,000 trials, about 3.2$\times$ the amount of real data trials from all participants. 
Figure~\ref{fig:synthesized_trials} illustrates how submovements from the synthesized and real data are distributed. 
As decomposition continues for a given trial its duration increases and the resulting submovements tend to have smaller displacement. 
These trends in our synthesized data were similar to real data. 
However, the synthesized data represents more trials with more overlapping submovements than the real data, possibly due to our lack of modeling of covariance between location, amplitude, and dilation. 
To avoid this, we rejection sampled our synthetic trials to have the same overlap ratio distribution as the real data, resulting in a synthetic data set of 6,406 trials.


\subsection{Results}
Similarly to the parameter refinement analysis in Section \ref{sec:weightedloss} above, we compared three different decomposition refinement methods: \textit{Initial guess}, \textit{MSE}, and \textit{Wavelet-weighted}. 
However, we now  evaluate them relative to known ground truth submovement counts from the synthetic data. 

\subsubsection{Submovement Counts}

\begin{figure}
    \centering
    \centering
    \includegraphics[width=\columnwidth]{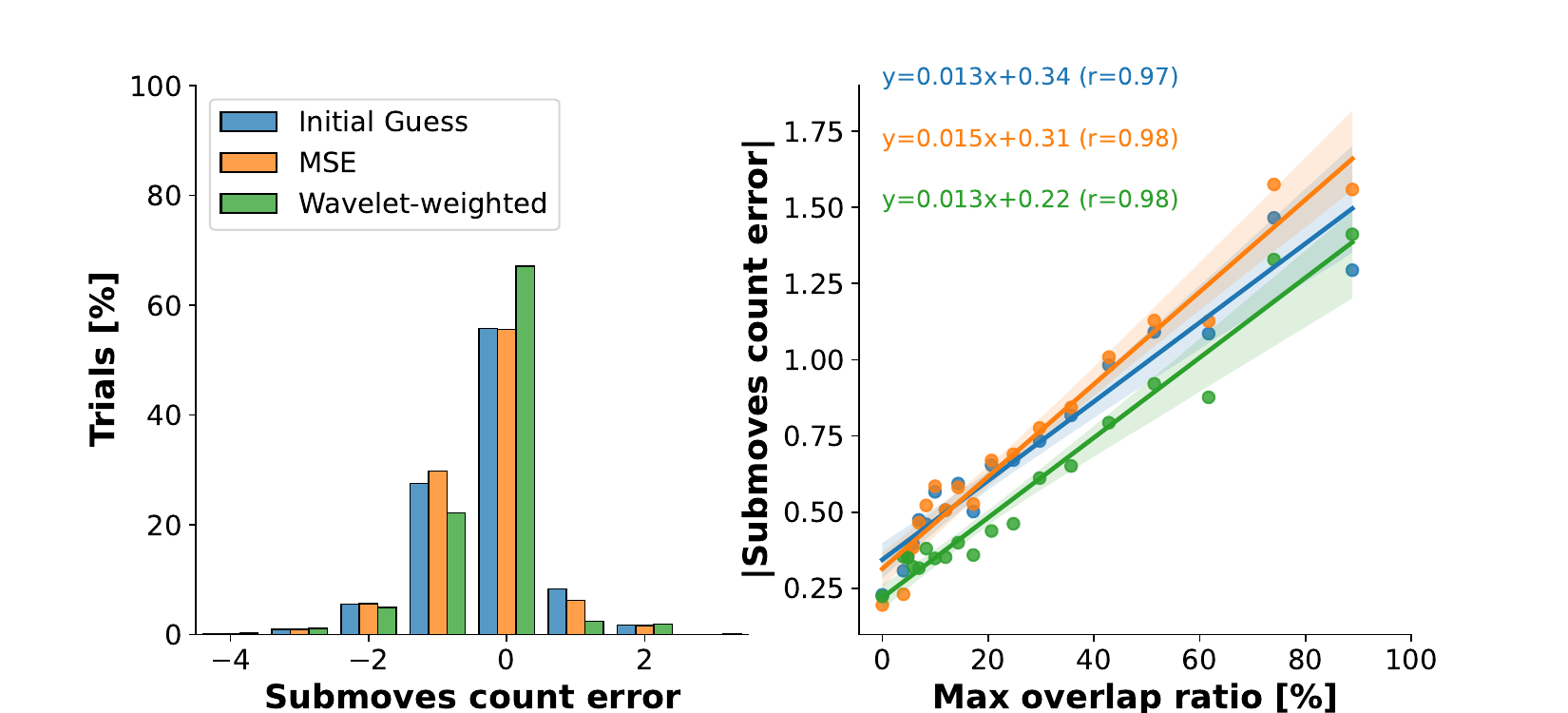}
    \vspace{-6mm}
    \caption{(Left) a comparison of different wavelet decomposition techniques over our synthetic data, note that wavelet-weighted loss improves the 0-error count and under-predicts submovement count less often than other methods. (Right) a scatter plot of overlap ratio vs submovement count error, demonstrating that highly overlapped submovements become linearly more difficult to detect.}
    \label{fig:synthesized_trial_count_errors}
\end{figure}

    
    
    
Results show that the \textit{wavelet-weighted} loss yields the most trials with correct submovement counts (67.124\%) compared to \textit{MSE} (55.589\%) and \textit{Initial guess} (55.713\%).
All three methods tend towards under-reporting submovement counts: \textit{Wavelet-weighted} (28.489\% vs. 4.387\%), \textit{MSE} (36.528\% vs. 7.883\%), and \textit{Initial guess} (34.140\% vs. 10.147\%).
This bias towards under-reporting submovement count is explainable.
As submovements overlap more substantially detecting them individually becomes more difficult, with two 100\% overlapped submovements of similar dilation often being completely indistinguishable from a single movement.
In fact we tune our algorithm's termination criteria specifically to avoid over-iterating and producing multiple small submovements of this type.


\subsubsection{Impacts of Overlap}
Figure~\ref{fig:synthesized_trial_count_errors} validates that higher overlap ratio yields higher submovement count errors, as expected. 
We perform linear regression for the (absolute value of) submovement count error by overlap ratio. To normalize the number of samples per bin, the submovement count error was averaged over logarithmic overlap ratio bins. 
Slopes of these count error vs overlap ratio lines were similar for all methods \textit{Wavelet-weighted} (0.0131), \textit{MSE} (0.0151), and \textit{Initial guess} (0.0130). However, the intercept of \textit{Wavelet-weighted} was 0.219 which is about two thirds of \textit{MSE} (0.314), and \textit{Initial guess} (0.344). This indicates that \textit{Wavelet-weighted} loss yields smaller submovement count error than the other metrics regardless of the amount of overlaps between wavelets.

\subsection{Comparison to Speed Threshold(s) and Persistence 1D}

As previously mentioned, common strategies for detecting non-overlapping submovements from (filtered) time series include single/dual threshold velocity segmentation \cite{walker1997program} or momentum-based segmentation such as persistence 1D \cite{lee2020autogain}.
All such segmentation techniques, by definition, delineate \emph{non-overlapping} submovements, declaring any given point in time to belong to exactly one submovement at most.
For this reason we expect that our approach will tend to report:

\begin{itemize}
    \item \emph{Higher} submovement counts than dual threshold detectors but \emph{lower} counts than persistence 1D
    \item \emph{Lower duration} submovements than alternate methods due to sensitivity to short submovements
    \item \emph{Lower displacement} submovements than alternate methods due to sensitivity to small submovements
\end{itemize}

\begin{figure}
    \centering
    \includegraphics[width=0.5\columnwidth]{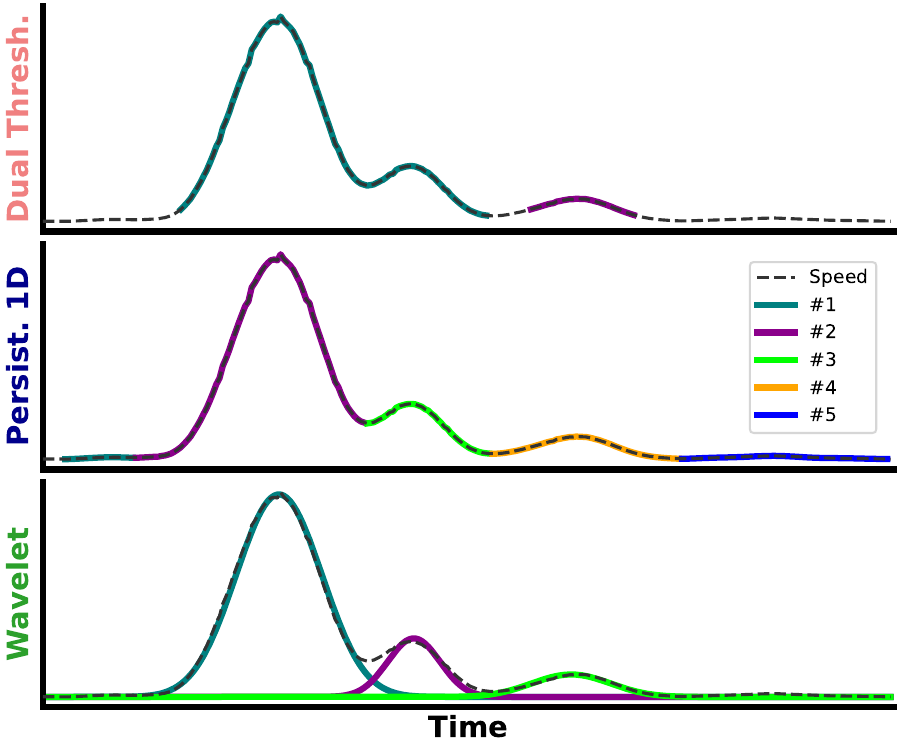}
    \caption{Qualitative comparison of the dual threshold, persistence 1D, and our wavelet detection algorithms. The dual threshold detector starts and stops submovements exactly at its thresholds, while persistence 1D identifies submovements based on changes in higher order derivatives. As a result the persistence 1D algorithm can segment overlapping submovements, but also \emph{completely} segments the time series, resulting in submovements being declared in periods of little-to-no motion. Our algorithm attempts to compromise between the two, selecting similar peaks but ignoring low motion periods. Note: The plots' y-axis crossings are drawn slightly below 0 to demonstrate this.}
    \label{fig:compare_example}
\end{figure}

Reporting higher submovement counts than speed-threshold techniques is intuitive as overlapped submovements count as multiple submovements instead of a single large and complex submovement with online correction.
Our application of persistence 1D considers periods of low motion as submovements and therefore artificially increases submovement count by completely segmenting the time series, increasing submovement counts significantly.
The dual threshold and persistence 1D methods reporting longer average duration submovements makes sense as their respective "merged" and added low-motion submovements bias them towards longer periods of time in a single submovement.
Since velocity-threshold techniques require a fixed start/stop speed threshold, they tend to miss long, slow submovements in favor of shorter, faster submovements, which are generally of larger displacement.
On the other hand, persistence 1D always segments long slow submovements, but over-approximates the displacement of them by including neighboring submovement overlap in segmentation. 
Our method avoids these issues as wavelet decomposition prioritizes the net \emph{displacement} of the movement, not necessarily its velocity or duration alone, treating long slow submovements as equally important to shorter, faster ones.

\begin{figure*}
    \includegraphics[width=\textwidth]{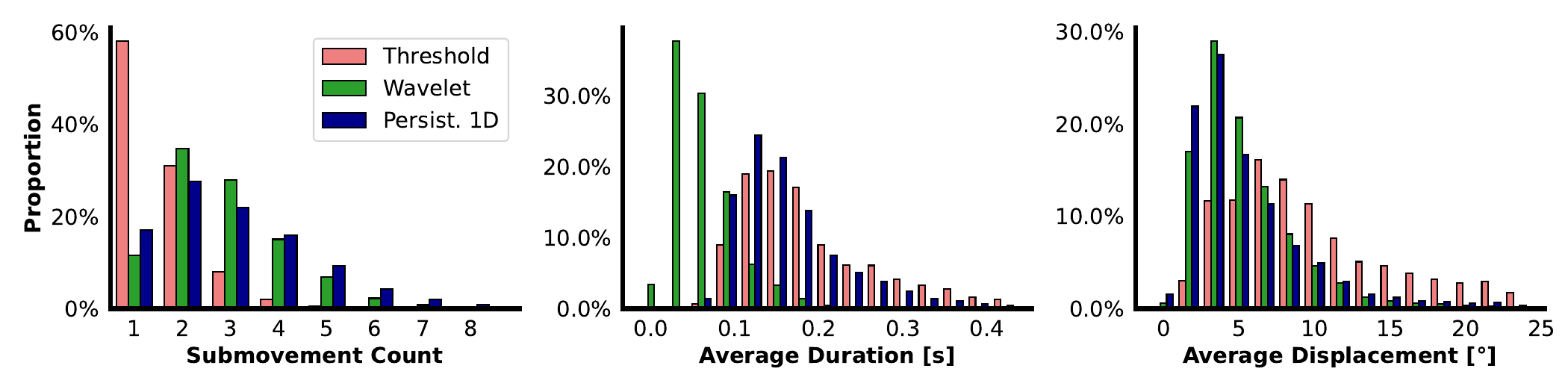}
    \vspace{-8mm}
    \caption{Comparison of our method (in green) to results from a dual-threshold detector with 4  $^{\circ}$/s start threshold and 8 $^{\circ}$/s stop threshold (light red) and persistence 1D with a threshold of 0.2 (in dark blue). (Left) detected submovement count shows we find more submovements than dual threshold but fewer than persistence 1D. (Center) submovement duration reveals that our detected movements are significantly (temporally) shorter on average than those detected by the dual-threshold and persistence 1D methods. (Right) average submovement displacement demonstrates that we tend to detect submovements of smaller displacement than the dual-threshold method but similar to persistence 1D.}
    \label{fig:dualthresh_comparison}
\end{figure*}

To validate the claims above we compare our decomposition with both dual threshold and persistence 1D speed segmentation over the real data from Section \ref{sec:realdata}.
We compare using real data, as the statistics and submovement model used to create our synthetic data set are based on our algorithm, and thus would likely bias towards our approach.
In all methods we apply the same Butterworth low-pass filter to the incoming (positional) motion data. 
We analyzed our real data using simple implementations of dual-threshold velocity segmentation \cite{boudaoud2022mouse} and persistence 1D \cite{persist1D}.
In the dual threshold detector, a submovement is declared from when mouse speed surpasses the start threshold (8 $\degree$/s) until the speed drops below the end threshold (4 $\degree$/s).
Following each submovement, a refractory period of 80 ms is enforced in which no new submovements are detected.
In the persistence 1D method we use a persistence threshold of 0.2 to segment submovements.
Figure \ref{fig:compare_example} provides an example of qualitative differences in the results provided by all 3 algorithms on a sample real data time series.

\subsubsection{Submovement Count Analysis}
Figure~\ref{fig:dualthresh_comparison} (left) illustrates the differences in submovement count distributions. 
We performed a Wilcoxon signed rank test to validate these differences on the average submovement count per user. 
Results showed a significant difference between the wavelet and threshold as well as the wavelet and persistence 1D method (z = 3.668, p < .001), but no significant difference between the wavelet and persistence 1D method. 
This supports the hypothesis that our detector reports higher submovement counts than dual threshold method, but fails to validate that our wavelet-based detector reports \emph{lower} submovement counts than the persistence 1D method.

\subsubsection{Submovement Duration Analysis}


Figure \ref{fig:dualthresh_comparison} (center) demonstrates differences in the average duration of submovements detected by the algorithms.
Note that the wavelet detector is able to detect submovements with peak amplitude below the dual threshold start threshold, so these static threshold criteria cannot be used to determine duration (they result in $\sim$20\% of detected submovements being ignored).
Instead we opt for a 1\% peak amplitude criteria to determine the start/end of a submovement and resulting duration.

As expected our wavelet method reports substantially shorter and more consistent duration submovements ($\mu_{w}$ = 76.21 ms, $\sigma_{w}$ = 38.61 ms) than the dual threshold detector ($\mu_{t}$ = 209.93 ms, $\sigma_{t}$ = 90.50 ms) and persistence 1D ($\mu_{p}$ = 179.60 ms, $\sigma_{p}$ = 73.56 ms).
Compared to our wavelet method, the dual threshold and persistence 1D detectors reported a 133.72 ms (3.46 $\sigma_{w}$) and 103.39 ms (2.68 $\sigma_{w}$) longer average duration respectively.
We performed a paired t-test on the average submovement duration per subject to validate these differences.
Results showed a significant difference (p < .001) between all 3 methods, with the largest effect between the wavelet and threshold methods ($t_{12}$ = 21.95), a similar difference between the wavelet and persistence 1D methods ($t_{12}$ = 18.63) and the smallest difference between the threshold and persistence 1D methods ($t_{12}$ = 8.78). 

\subsubsection{Submovement Displacement Analysis}


Note that it is difficult to fairly compare these methods in displacement of submovements as many of the movements detected by our wavelet method do not meet the start threshold for the dual-threshold detector or extend well beyond the bounds of persistence 1D.
Our method reports smaller displacement submovements on average than the dual threshold detector ($\mu_{w}$ = 6.13$^{\circ}$, $\sigma_{w}$ = 3.46$^\circ$ vs $\mu_{t}$ = 10.38$^{\circ}$, $\sigma_{t}$ = 5.36$^{\circ}$) but these differences in displacement are not nearly as large as those between reported durations.
The wavelet detector and persistence 1D report nearly identical statistics ($\mu_{p}$ = 6.19, $\sigma_{p}$ = 4.10$^{\circ}$).
We performed a paired t-test over the average submovement displacement per subject. 
Results showed a significant difference between the wavelet and dual threshold methods ($t_{12}$ = 14.20, p < .001) as well as the dual threshold and persistence 1D methods ($t_{12}$ = 20.95, p < .001), but not between our wavelet detector and the persistence 1D method.
The distribution in Figure \ref{fig:dualthresh_comparison} (right) demonstrates that much of the resulting difference in mean is likely driven by the (slight) bias of the dual threshold and persistence 1D detectors towards large(r) single-submovement displacements.

These results support our hypothesis that wavelet methods report significantly lower average displacement submovements than the dual threshold detector, but reject the hypothesis that wavelet detectors report significantly smaller submovements than persistence 1D.
We postulate that we do, in fact, often predict smaller submovements than persistence 1D, but that the average displacements for the persistence 1D method are artificially decreased by the low-to-no motion windows that the algorithm considers as submovements.

\section{Discussion}
\subsection{Limitations and Suggested Improvements}

\subsubsection{CWT Optimizations}
The 10$\sigma$ (width) length assumption of SciPy's \texttt{scipy.signal.cwt} method over-represents submovement duration.
We prefer to post-process fit wavelets to identify the significant portion of motion (i.e., 99\% displacement) and use this to report the movement duration, as is done in the analysis of Figure \ref{fig:dualthresh_comparison}.
Additionally, \texttt{scipy.signal.cwt} requires the width ($\sigma$) parameter to be provided as an integer value.
We resolve this by dividing the provided $\sigma$ value by 100 inside of our submovement model, effectively stepping the dilation in increments of 0.01.
It is possible this step could create problematic quantization effects in some cases, and should be considered for those trying to very accurately fit initial guesses for wavelets, at the expense of compute cost.

\subsubsection{Challenging Conditions}

\begin{figure}
    \centering
    \includegraphics[width=0.49\columnwidth]{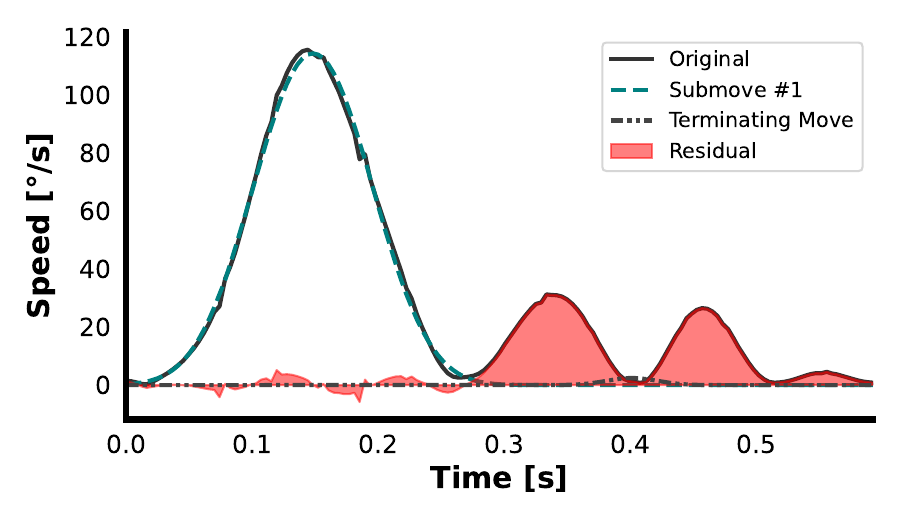}
    \includegraphics[width=0.49\columnwidth]{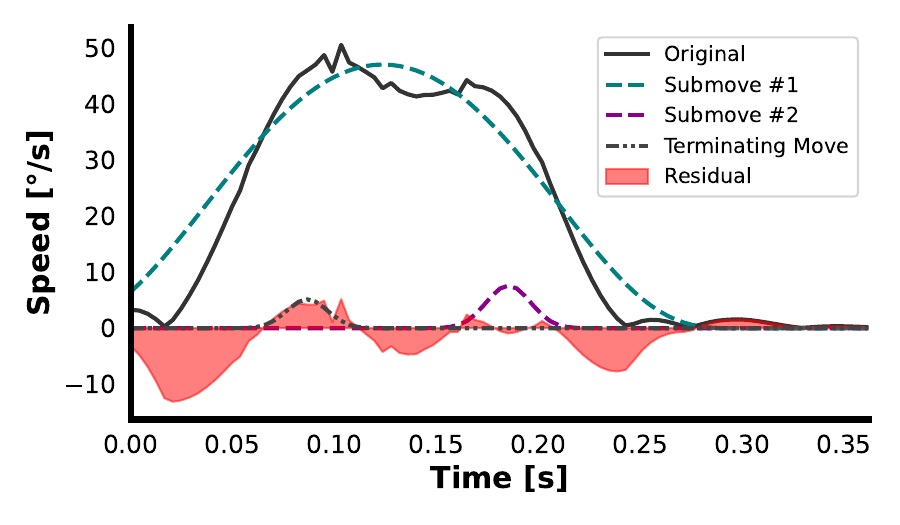}
    \caption{(Left) An example of an early termination caused by a misfit submovement. In this case the submovement in gray is fit to the center of two submovements and fails to move towards one peak or the other with regression. The result is terminating early, leaving a large residual. (Right) An example of an over estimated/poorly fit first submovement width impacting the following submovement decomposition, in this case underestimating their amplitude.}
    \label{fig:failures}
\end{figure}


One failure mode for our termination criteria involves misfit submovements which happen to be small in amplitude.
When an early submovement is fit to a small amplitude through either poor initial CWT estimation or a failure of our wavelet-weighted linear regression, the result is a small submovement being predicted \emph{before} larger submovements.
This can result in early termination as shown in Figure \ref{fig:failures} (left).

An additional failure mode involves over estimating the width ($\sigma$) parameter for a large submovement in the trial.
Our wavelet-weighed loss metric is intended to avoid this case, but can still fail in cases where the overlap of the underlying submovement is such that the loss is still relatively small.
Figure \ref{fig:failures} (right) provides an example of this failure mode, with Submove \#1 fitting two submovements and the resulting 2nd and terminating submovement modeling the residual peaks.
This can be worked around by either limiting the values of $\sigma$ being searched or by building in a larger penalty for mismatching near the peak of the wavelet in the loss metric.




\subsubsection{Alternative Termination Criteria}
Additional decomposition loop termination criteria can, and should, be investigated depending on the use of the decomposition strategy.
Researchers interested in studying large primary motions used to approach a target may use a more aggressive termination criteria than those studying the iterative motions used to "zero in" on small targets.

One alternative is thresholding the \emph{displacement portion}, or distance traveled in the current submovement divided by the net displacement of the time series of interest.
This criteria works well in short time series with low submovement count, but struggles to successfully segment small submovements from larger time series.
A similar option is using the \emph{net (positive) residual} at each decomposition step to terminate, effectively thresholding remaining displacement unexplained by fit submovements.
By considering the positive residual only motion originally present in the time series, not that caused by misfit wavelets, is counted towards termination.
This methods can lead to very high submovement counts when the algorithm encounters a time series it struggles to decompose.
A final option is to threshold the \emph{displacement} of the current detected submovement itself.
This method offers the intuitive advantage of stopping decomposition once movements become "small enough" that they no longer contribute significantly to overall motion.
Since typically CWT decomposition finds the largest submovements first, this method makes sense; however, our added refitting step can significantly reduce the amplitude of submovements, resulting in early termination in some cases.

\subsubsection{Confidence Metrics}

The (integral of the) residual at the end of decomposition could be used as a confidence metric indicating the quality of fit.
We suggest evaluating the net positive/negative residuals independently as part of the confidence metric to: (positive) describe how much motion is still unfit by the algorithm/detect early terminations and (negative) assess quality of fit of each submovement and find over-estimated widths that may have impacted the quality of other submovement fits.

\section{Conclusions}
We introduce a new wavelet inspired method for decomposing overlapping submovements from (mouse) input time series.
Our method uses a CWT decomposition of a one dimensional speed (velocity magnitude) time series as an initial guess for a least-squares based refinement step using a wavelet-weighted loss metric.
By applying this refinement we demonstrate avoiding fitting overly broad wavelet dilations to multiple submovements and significantly decrease errors in the location of submovement speed peaks, where we move fastest.
Using real data, we demonstrate that our method accurately detects overlapping movements and reduces both RMSE and unresolved displacement (residuals) compared to a naive CWT or MSE-based refinement step alone.
Using synthetic data we demonstrate our wavelet-weighted loss refinement step improves submovement count error and demonstrates more robustness against overlap than a naive CWT or RMSE-based refinement.
We compare our method to a dual threshold detector and persistence 1D and demonstrate significant differences from both of them.
We suggest our technique to researchers interested in studying the \emph{overlap} or online correction of submovements as it avoids the issue of non-overlapping time-segmentation techniques such as dual-threshold detectors and persistence 1D.
Finally we discuss remaining limitations of our method and suggest possible additional improvements.
We hope that submovement decomposition and analysis benefits from our work and continues to be applied to broader and more varied areas of study!